\documentclass[10pt,conference,english,twocolumn]{IEEEtran}
\IEEEoverridecommandlockouts
\usepackage{babel}
\usepackage[babel]{microtype}
\usepackage[babel]{csquotes}

\usepackage[T1]{fontenc}

\usepackage[svgnames]{xcolor}
\usepackage{graphicx}

\usepackage{tabularx}
\usepackage{booktabs}

\usepackage{amsmath}
\usepackage{amsfonts}
\usepackage{amssymb}
\usepackage{amsthm}
\usepackage{bm}
\usepackage{siunitx}
\usepackage{mathtools}
\usepackage{dsfont}
\newtheorem{defn}{Definition}

\usepackage[math]{blindtext}

\usepackage{cite}
\usepackage{titlesec}

\usepackage[moderate]{savetrees}

\begin{document}

\title{{Multi-Connectivity for UAVs: A Measurement Study of Integrating Cellular, Aerial Mesh, and LEO Satellite Links}
\thanks{This work was supported in part by the CELTIC-NEXT Projects 6G for Connected Sky (6G-SKY, C2021/1-9) and 3D NETworks for 6G Mobile Communications Applications (3D-NET, C2025/1-11), both funded by the Swedish Innovation Agency (Vinnova), as well as by the SSF Research Centre Sustainable Mobile Autonomous and Resilient 6G SatCom (SMART-6GSAT, CSG23-0001).}
}

\author{Aygün~Baltaci\textsuperscript{\S},
    Irshad A. Meer\textsuperscript{*},
    Mustafa~Ozger\textsuperscript{$\ddagger$*}, 
    Cicek Cavdar\textsuperscript{*}, and 
    Dominic~Schupke\textsuperscript{\S}
   
		\\
		\textsuperscript{*}KTH Royal Institute of Technology, Sweden \;\;
  	\textsuperscript{$\ddagger$}Aalborg University, Denmark 
    \\
    \textsuperscript{$\S$}Airbus Central Research and Technology, Germany 
		\\Email: \textsuperscript{*}\{iameer, cavdar\}@kth.se,  \textsuperscript{$\ddagger$}mozger@es.aau.dk, \textsuperscript{\S}\{ayguen.baltaci, dominic.schupke\}@airbus.com	}

\maketitle
\thispagestyle{empty}
\begin{abstract}\noindent\boldmath
Future uncrewed aerial vehicle (UAV) systems increasingly combine heterogeneous communication technologies, such as low-latency aerial mesh, terrestrial cellular, and satellite links, to improve robustness and coverage. Multipath transport is a natural mechanism for aggregating these links, yet its ability to support real-time UAV services in highly heterogeneous environments remains insufficiently characterized.
We present a measurement-driven study based on UAV flight experiments in an integrated network comprising UAV-to-UAV aerial mesh, private cellular, and low Earth orbit (LEO) satellite connectivity. 
Using Multipath TCP (MPTCP) as a representative lossless, in-order multipath transport framework, we find that aggregation can preserve end-to-end connectivity under severe link outages. However, large round-trip time (RTT) heterogeneity amplifies packet reordering, leading to substantial receiver-side buffering and bursty delivery. In addition, when the available links do not provide sufficient capacity for the offered load, pronounced sender-side buffering emerges. These effects cause real-time streaming to violate delay constraints, including cases where aggregate capacity is sufficient.
To interpret these results, we formalize the distinction between \emph{connectivity continuity} and \emph{service continuity} and show empirically that maintaining connectivity is necessary but not sufficient for timely real-time delivery in multi-technology UAV networks. The findings motivate multipath designs that explicitly account for delay constraints, rather than optimizing for connectivity alone.

\end{abstract}

\section{Introduction}
\label{sec:introduction}


Uncrewed aerial vehicles (UAVs) are deployed in missions that demand both high reliability and strict real-time performance, including remote piloting, infrastructure inspection, and emergency response \cite{baltaci2021survey}. Achieving these requirements is challenging due to intermittent coverage, mobility-induced disruptions, and harsh propagation conditions \cite{meer2024mobility}. To improve robustness, UAV systems increasingly rely on the concurrent use of multiple communication technologies, such as low-latency UAV-to-UAV aerial mesh links, terrestrial cellular networks, and non-terrestrial satellite connectivity \cite{vondra2018integration}. This multi-technology integration is widely viewed as a key enabler for resilient UAV communications \cite{salehi2023reliability}.

A common approach to leveraging such heterogeneous links is multipath transport, in particular multipath Transmission Control Protocol (MPTCP), which distributes traffic across multiple paths and can preserve end-to-end connectivity under individual link failures.
In particular, lossless, in-order multipath mechanisms are often assumed to provide reliability by maintaining communication as long as at least one path remains operational \cite{BaltaciIEEEAccess2023AerialMP}. This perspective has motivated the use of multipath transport as a building block for UAV deployments that span terrestrial and non-terrestrial segments.

However, most prior work on multipath transport has focused on throughput, fairness, and aggregate capacity utilization, while real-time requirements are less explored. Many UAV applications are latency-sensitive and require delay bounds to be satisfied continuously to support safe and effective operation. In integrated UAV networks, links can differ substantially in both latency and available capacity, raising the question of whether conventional lossless, in-order multipath transport can provide not only connectivity but also timely delivery for latency-sensitive traffic.


In this paper, we address this question through a measurement-driven study based on UAV flight experiments in an integrated network comprising three fundamentally different technologies: a low-latency aerial mesh link, a private terrestrial cellular link, and a low Earth orbit (LEO) satellite link. We aggregate these links using a lossless, in-order multipath transport mechanism, enabling the UAV to maintain end-to-end connectivity under severe link outages. Our measurements reveal two practical limitations. First, when the available links do not provide sufficient capacity for the offered load, multipath transport can lead to pronounced sender-side buffering. Second, when the aggregated paths exhibit large round-trip time (RTT) heterogeneity, strict in-order delivery amplifies reordering, causing substantial receiver-side buffering and bursty packet delivery. As a result, latency-sensitive traffic can violate delay constraints even when aggregate capacity is sufficient.

\begin{figure*}[t]
    \centering
    \includegraphics[width=0.88\textwidth]{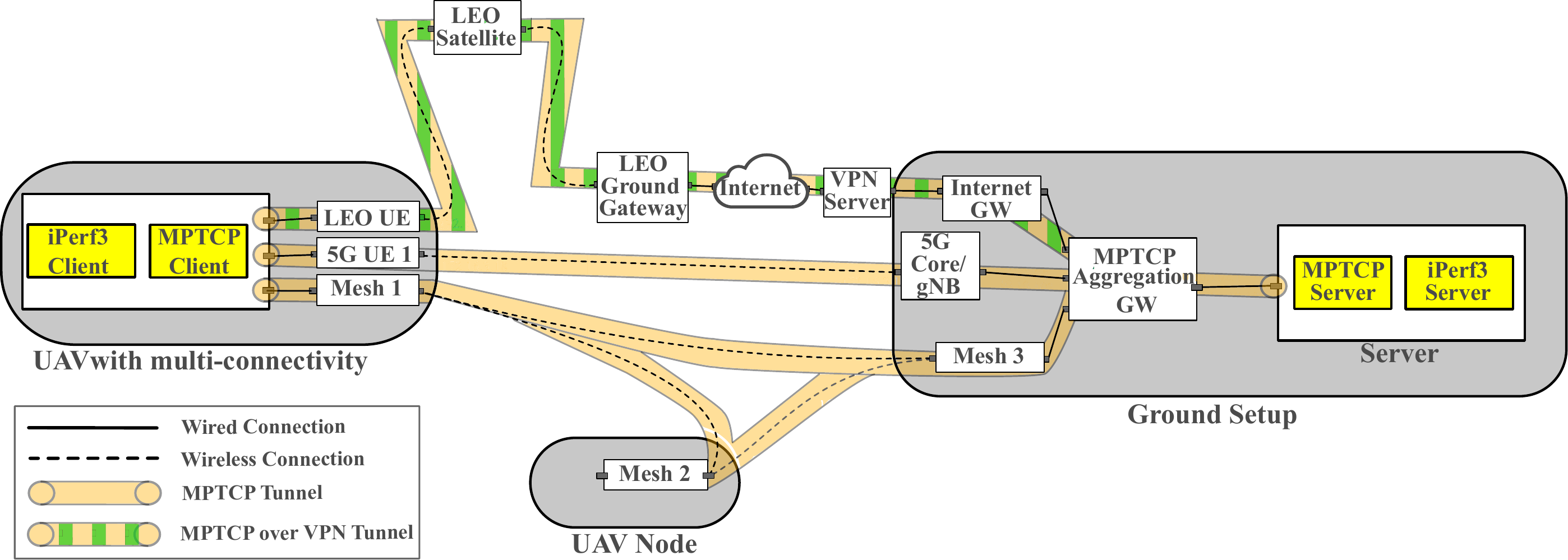}
    \caption{Multi-technology link handover demonstration setup at the Airbus test site in Ottobrunn/Munich, Germany.}
    \label{fig:demo_setup}
    \vspace{-1em}
\end{figure*}


To interpret these results, we distinguish between two notions that are often conflated in robust UAV communication design: \emph{connectivity continuity} and \emph{service continuity}. Connectivity continuity captures the ability to maintain an end-to-end communication path despite link failures. Service continuity captures the ability to sustain application operation by delivering data within the required delay bounds under the offered load, which implicitly depends on both latency and available capacity. Using formal definitions and empirical evidence, we show that in heterogeneous multi-technology UAV networks, connectivity continuity is necessary but does not imply service continuity. In particular, lossless, in-order multipath transport can amplify delay asymmetries across paths, leading to buffer growth and transient latency spikes that break real-time service guarantees.

This paper focuses on characterizing the interaction between lossless multipath transport and heterogeneous path conditions in realistic UAV deployments. As such, we do not attempt an exhaustive comparison of individual communication technologies, nor do we evaluate alternative transport protocols.


The main contributions of this paper are as follows:
\begin{itemize}
\item We present a flight-validated measurement study of multipath transport in a multi-technology UAV network integrating aerial mesh, ground-to-air cellular, and LEO satellite links, and characterize the impact of heterogeneous path conditions on real-time traffic.
\item We show that while lossless multipath transport preserves end-to-end connectivity under severe link outages, it can induce sender-side buffering under capacity shortfalls and receiver-side buffering with bursty delivery under large RTT heterogeneity, leading to violations of real-time delay constraints.
\item We formalize the distinction between connectivity continuity and service continuity, and provide empirical evidence that the former is a necessary but insufficient condition for the latter in heterogeneous terrestrial--non-terrestrial UAV networks.
\end{itemize}

\section{System Overview and Measurement Setup}
\label{sec:system}

This section describes the multi-technology communication system for UAV connectivity considered in this study and the corresponding measurement methodology. The objective is to characterize the behavior of lossless multipath transport under realistic terrestrial--non-terrestrial integration, rather than to optimize or redesign individual network components. Accordingly, we focus on system aspects that directly influence end-to-end delay, receiver buffering, and service continuity.

\subsection{Multi-Technology UAV Communication Architecture}
The UAV is equipped with multiple heterogeneous wireless interfaces, enabling concurrent
connectivity across different access technologies and exploiting their complementary characteristics
to support end-to-end data transmission under varying coverage conditions. The end-to-end
multi-technology communication setup used in the flight measurement at the Airbus test site in Ottobrunn/Munich, Germany, is illustrated in Fig.~\ref{fig:demo_setup}. Specifically, the UAV can establish
connectivity through the following classes of communication links:

\begin{itemize}
\item \emph{Aerial mesh link}: A low-latency air-to-air (A2A) communication link connecting the UAV to nearby airborne nodes forming a wireless mesh network. This link provides short-range connectivity with low propagation delay and is typically available under line-of-sight conditions. In Fig.~\ref{fig:demo_setup}, this link connects the Mesh 1 and Mesh 2 nodes. A2A communication is realized using a proprietary mesh protocol implemented over Wi-Fi technology. In the considered setup, a mesh-capable ground gateway node (Mesh 3 in Fig.~\ref{fig:demo_setup}) receives data from the UAV of interest (UAV with multi-connectivity, i.e., Mesh 1, in Fig.~\ref{fig:demo_setup}) whenever it is within coverage. An additional airborne mesh node (Mesh 2 Fig.~\ref{fig:demo_setup}) can serve as a relay, extending connectivity when the UAV moves beyond the ground gateway’s direct communication range.

\item \emph{Terrestrial cellular link}: 
A private 5G link provides ground-based coverage for the UAV as an aerial user. The link is implemented using a private 5G network with a ground-deployed femtocell operating at 3.7~GHz. Due to the femtocell’s limited transmit power, coverage is confined to approximately $<100$~m. The 5G core network is connected to the aggregation gateway. Compared to the aerial mesh link, the air-to-ground cellular link (between 5G UE 1 and the ground gNB in Fig.~\ref{fig:demo_setup}) provides moderate latency and data rates, and it supports mobility. 
Its performance and availability may be affected by UAV altitude, cell boundary crossings, antenna tilt, and time-varying radio propagation conditions.

\item \emph{Non-terrestrial link}: A satellite-based connection providing beyond-line-of-sight connectivity. This link ensures wide-area coverage even in the absence of terrestrial infrastructure, but exhibits substantially higher round-trip time (RTT) than both aerial mesh and terrestrial cellular links primarily due to satellite backhaul. Its availability depends on satellite visibility, link scheduling, and backhaul conditions.
\end{itemize}

The UAV maintains concurrent connectivity over the available communication links. The data traffic may be distributed across multiple paths by the transport layer when multiple links are available. At the receiver side, traffic from the heterogeneous links is terminated at a common gateway and forwarded to the application endpoint. 


\subsection{Transport Layer Mechanism and Traffic Characteristics}

Traffic aggregation across heterogeneous links is performed using a lossless, in-order multipath transport mechanism, which is multipath Transmission Control Protocol (MPTCP) \cite{ford2013tcp}. This design reflects current practice in multipath-capable transport protocols, which aim to improve robustness by maintaining multiple transport subflows while guaranteeing reliable and ordered delivery.

The UAV generates a constant-rate real-time traffic stream with a source rate of 10 Mbit/s, which remains fixed throughout the experiment.
Packets are scheduled across the available paths according to transport-layer decisions. To support the MPTCP experiments, we configured the transport layer accordingly. In particular, we enabled the lowest-RTT scheduler to prioritize the path with the smallest round-trip time and thereby minimize end-to-end latency. We also configured the MPTCP congestion control, which is a core transport-layer function alongside data scheduling. Unless otherwise stated, we used TCP Reno congestion control algorithm, a standard TCP variant that reacts to packet loss by reducing (halving) the congestion window.
At the receiver, MPTCP enforces strict in-order delivery: packets that arrive out of order are buffered until the missing data is received, and data is released to the application only in sequence. As a result, end-to-end packet age and service continuity are directly influenced by path delay heterogeneity and receiver-side buffering dynamics.
We do not employ any application-layer adaptation, packet dropping, or deadline-aware scheduling. This design choice isolates the effects of heterogeneous path delays and the lossless, in-order delivery behavior of MPTCP on the timeliness of real-time traffic.
We adopt MPTCP because it is a mature and practically deployable multipath transport protocol that provides reliable, in-order delivery across multiple subflows. Our goal is not to optimize the transport layer, but to use a representative lossless multipath mechanism to expose the effects of heterogeneous path delays and capacities.

\begin{figure}[t]
    \centering
    \includegraphics[width=\linewidth]{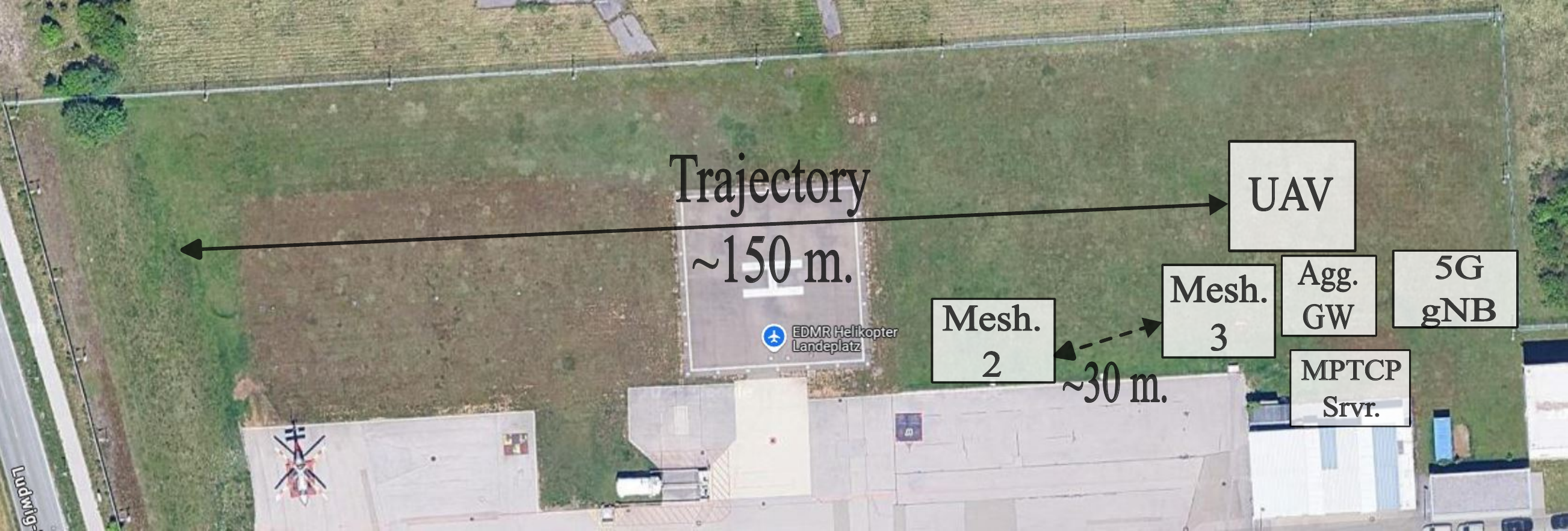}
    \caption{Flight measurement setup.}
    \label{fig:measurement_setup}
    \vspace{-3mm}
\end{figure}

\subsection{Measurement Methodology}

Measurements are conducted during real UAV flight experiments as shown in Fig. \ref{fig:measurement_setup} in which the availability and quality of the communication technologies vary over time. The UAV was flown horizontally at altitudes up to $30$ m to fly away from the 5G base station and the Mesh UEs. The experiments include operating regimes where one or more links become unavailable due to mobility, coverage limitations, or propagation conditions, while at least one link remains operational.

During each experiment, we record the following metrics at the receiver:
\begin{itemize}
\item Per-path RTT measurements obtained at the transport layer,
\item Traffic volume transmitted over each path,
\item Receiver-side buffer occupancy over time,
\item Packet arrival times and instantaneous application-layer receive rates.
\end{itemize}

The resulting traces are used to directly analyze buffering dynamics and service continuity violations, and to perform trace-driven post-processing that correlates these outcomes with path delay heterogeneity.

\subsection{Scope and Limitations}

This paper focuses on characterizing the interaction between lossless multipath transport and heterogeneous path delays in realistic UAV deployments. As such, we do not attempt an exhaustive comparison of individual communication technologies, nor do we evaluate alternative transport protocols. 
Accordingly, we do not explicitly model wireless channels, mobility patterns, or traffic dynamics beyond what is captured in the measurements. The analysis centers on the implications of measured heterogeneous path conditions and path availability for service continuity under multipath aggregation.

\section{Connectivity vs. Service Continuity}
\label{sec:continuity}

UAV networks  with multi-technology are commonly designed to maximize robustness by leveraging multiple communication paths. In this section, we show that such robustness must be interpreted carefully for real-time UAV services by distinguishing between \emph{connectivity continuity} and \emph{service continuity}. While these notions often coincide in homogeneous networks, they diverge fundamentally under heterogeneous delay conditions, as encountered in integrated terrestrial--non-terrestrial deployments.

We consider a UAV transmitting a real-time data stream to a ground-based receiver through a set of $K$ parallel communication paths, denoted by
\begin{equation}
\mathcal{P} = \{1, 2, \dots, K\}.
\end{equation}
Each path $k \in \mathcal{P}$ is characterized by a time-varying capacity $C_k(t)$, an RTT $R_k(t)$, and an availability indicator $a_k(t) \in \{0,1\}$, where $a_k(t)=1$ indicates that path $k$ is operational at time $t$.

The UAV generates packets at a constant rate $r$ and transmits them using a lossless, in-order multipath transport mechanism. Packets arriving out of order at the receiver are buffered until missing packets are received, after which buffered packets are released for consumption by the application.

\subsection{Connectivity Continuity}

We first formalize the notion of connectivity commonly targeted by multipath transport protocols.

\begin{defn}[Connectivity Continuity]
The system maintains \emph{connectivity continuity} over an interval $[0,T]$ if, for all $t \in [0,T]$, at least one communication path is available, i.e.,
\begin{equation}
\exists k \in \mathcal{P} \;\text{s.t.}\; a_k(t) = 1.
\end{equation}
\end{defn}

Connectivity continuity ensures that packets can always be transmitted over a path and that end-to-end communication is not completely disrupted by individual link failures. 
The primary objective of our multipath transport mechanism for real-time UAV communication is to tolerate intermittent path availability by maintaining at least one end-to-end connection across multiple links spanning different communication technologies.

\subsection{Service Continuity}

For real-time UAV applications, requiring packets to be delivered within strict delay bounds in order to maintain safe and stable operation, connectivity alone is insufficient.
Let $\Delta(t)$ denote the end-to-end packet age at time $t$, defined as the elapsed time between packet generation at the UAV and packet delivery to the application at the receiver.

\begin{defn}[Service Continuity]
The system maintains \emph{service continuity} over an interval $[0,T]$ if the packet age satisfies
\begin{equation}
\Delta(t) \leq \Delta_{\max}, \quad \forall t \in [0,T],
\end{equation}
where $\Delta_{\max}$ is a service-specific delay bound.
\end{defn}

Service continuity captures the timeliness requirements of latency-sensitive UAV services and directly reflects the quality perceived by the application.

\vspace{-1mm}
\subsection{Divergence Under Heterogeneous Paths}


In networks where all paths exhibit comparable delays and the aggregate available capacity is sufficient to sustain the application data rate, connectivity continuity often aligns with service continuity: packets arrive with limited reordering, receiver buffering remains modest, and timely delivery is largely preserved. However, service continuity may degrade under two distinct conditions. First, when the aggregate available capacity
\begin{equation}
C_{\text{agg}}(t) = \sum_{k \in \mathcal{P}} a_k(t) C_k(t)
\end{equation}
falls below the source rate $r$, sender-side buffering accumulates even if path delays are similar. Second, when heterogeneous path delays or delay variability are present, strict in-order delivery can amplify reordering effects and induce receiver-side buffering.

Consider two paths $i,j \in \mathcal{P}$ with $R_i(t) \ll R_j(t)$ over a given time interval. Packets transmitted over the lower-latency path $i$ may arrive significantly earlier than packets transmitted over path $j$, causing the receiver to buffer early arrivals while waiting for delayed packets to preserve in-order delivery. Under a constant source rate $r$, the receiver buffer occupancy $B(t)$ grows proportionally to the RTT disparity, i.e.,
\begin{equation}
B(t) \sim r \cdot \bigl(R_j(t) - R_i(t)\bigr),
\end{equation}
until delayed packets arrive and buffered packets are released in a burst.
Such buffer growth and bursty release can lead to large transient packet ages and violate the service continuity condition, even when connectivity continuity is preserved and aggregate path capacity is sufficient.

\section{Measurement Results: Multipath Transport under Heterogeneous Links}
\label{sec:measurements}

This section presents measurement results obtained from real UAV flight experiments conducted as part of real UAV flight experiments on multi-technology network integration. All results are derived from packet captures collected at the MPTCP aggregation gateway and from transport-layer RTT estimates encoded in MPTCP packet headers. The objective is to characterize end-to-end behavior under heterogeneous link conditions, rather than to benchmark individual technologies in isolation.

\subsection{Individual Link Performance and Capacity Constraints}

We first summarize the performance characteristics of the individual communication links based on static ground measurements conducted prior to the flight experiments. Table~\ref{tab:link_performance} reports the achievable data rates, RTTs, and connection dynamics of the aerial mesh, private cellular, and satellite links. The aerial mesh link achieves data rates exceeding $30$~Mbit/s with very low RTT, making it capable of sustaining the $10$~Mbit/s real-time application used in the experiment. In contrast, both the private cellular and satellite links are limited to approximately $5$~Mbit/s under the tested conditions. Consequently, when the aerial mesh link is unavailable, the remaining links are individually insufficient to support the application source rate without inducing buffering at the transport layer. 

\begin{table}[t]
\centering
\caption{Individual link performance during static ground measurements.}
\label{tab:link_performance}
\renewcommand{\arraystretch}{1.2}
\begin{tabular}{lccc}
\toprule
\textbf{Metric} & \textbf{Aerial mesh} & \textbf{Private 5G} & \textbf{Satellite} \\
\midrule
Achievable data rate & $>$$30$~Mbit/s & $\sim$$5$~Mbit/s & $\sim$$5$~Mbit/s \\
Latency (RTT) & $\sim$$5$~ms & $\sim$$30$~ms & $150$--$200$~ms \\
Initial connection time & Few seconds & Few seconds & Few minutes \\
Reconnection time & Few seconds & Few seconds & Few minutes \\
\bottomrule
\end{tabular}
\end{table}

\subsection{Round-Trip Time Heterogeneity}

\begin{figure}[t]
    \centering
    \includegraphics[width=0.9\linewidth]{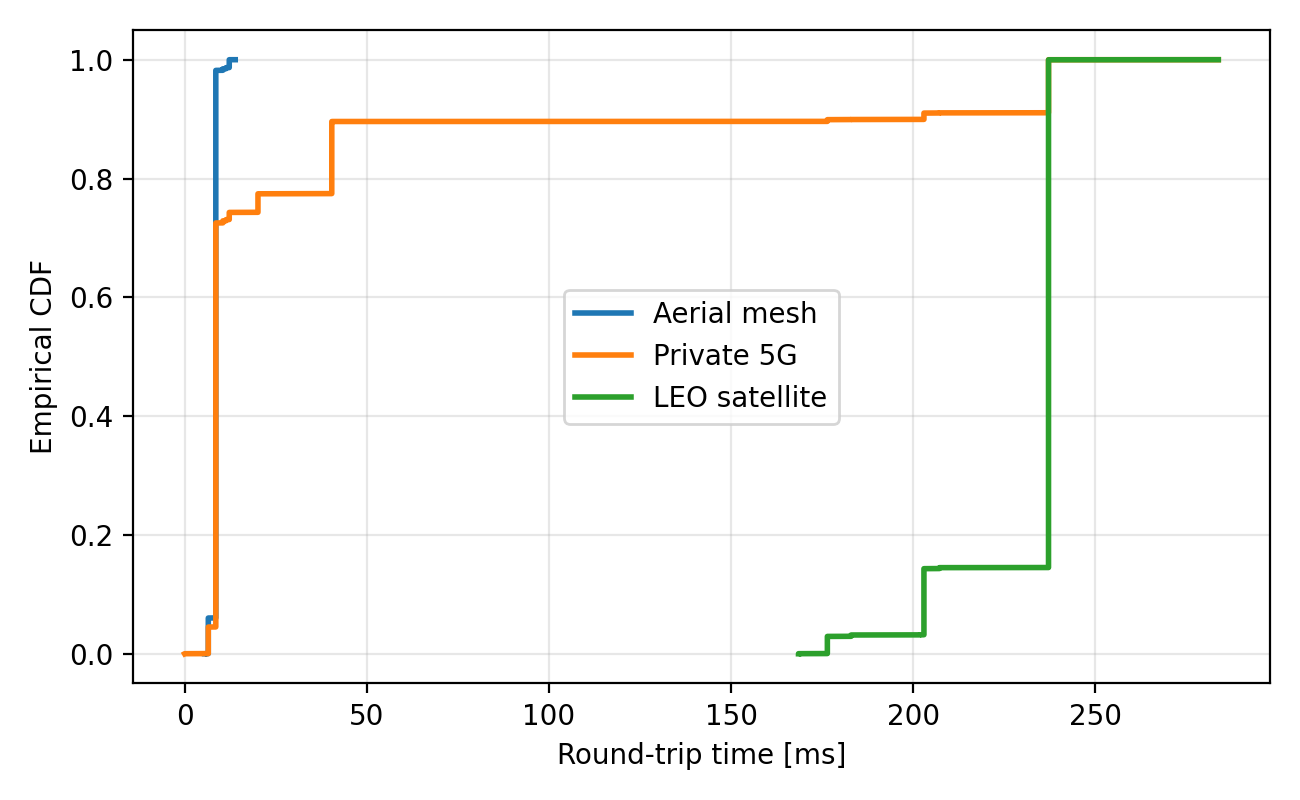}
    \caption{Empirical cumulative distribution function (CDF) of round-trip times (RTTs) observed
    at the transport layer during the UAV flight. RTT samples are derived from TCP RTT estimates
    recorded in MPTCP packet headers.}
    \label{fig:rtt_cdf}
    \vspace{-3mm}
\end{figure}

The empirical cumulative distribution function (CDF) of RTTs observed during the UAV flight is shown in Figure~\ref{fig:rtt_cdf}. RTT samples are derived from TCP RTT estimates recorded in MPTCP packet headers.
The aerial mesh link exhibits consistently low RTTs, typically on the order of a few milliseconds. The private cellular link operates at higher RTTs with increased variability, while the satellite link introduces substantially larger RTTs, with values exceeding $150$~ms for the majority of samples. These measurements confirm pronounced RTT heterogeneity across the available links, with delay differences exceeding one order of magnitude, directly impacting packet ordering and buffering behavior under lossless multipath transport.

\subsection{Cumulative Data Delivery and Link Dominance}

\begin{figure}[t]
    \centering
    \includegraphics[width=\linewidth]{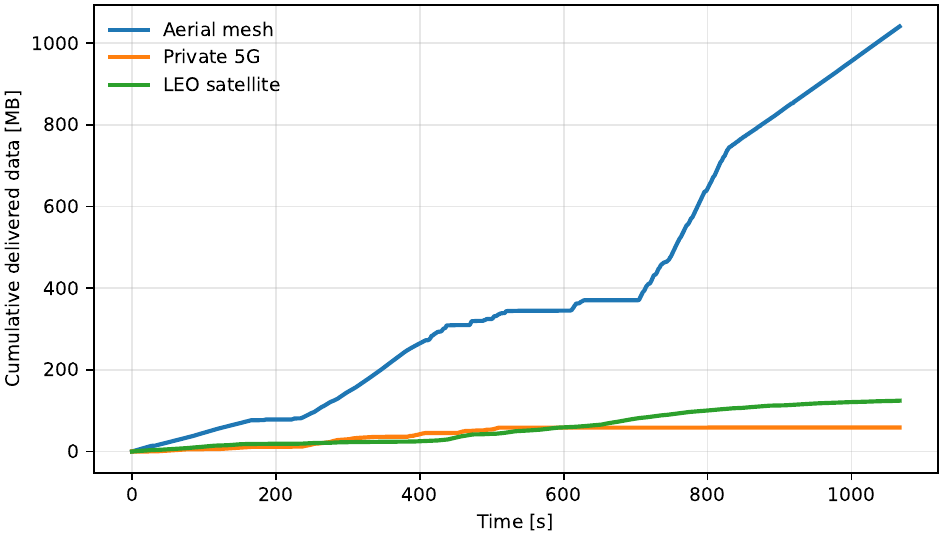}
    \caption{Cumulative data delivered over each communication link during the UAV flight.}
    \label{fig:cumulative_data}
\end{figure}

The cumulative amount of data delivered over each communication link during the flight is shown in Figure~\ref{fig:cumulative_data}. The aerial mesh link carries the majority of the transmitted data, reflecting its lower RTT and higher effective capacity. The private cellular and satellite links contribute smaller fractions of the total data volume and primarily serve as fallback paths when the mesh link becomes unavailable.

\begin{figure}[t]
    \centering
    \includegraphics[width=\linewidth]{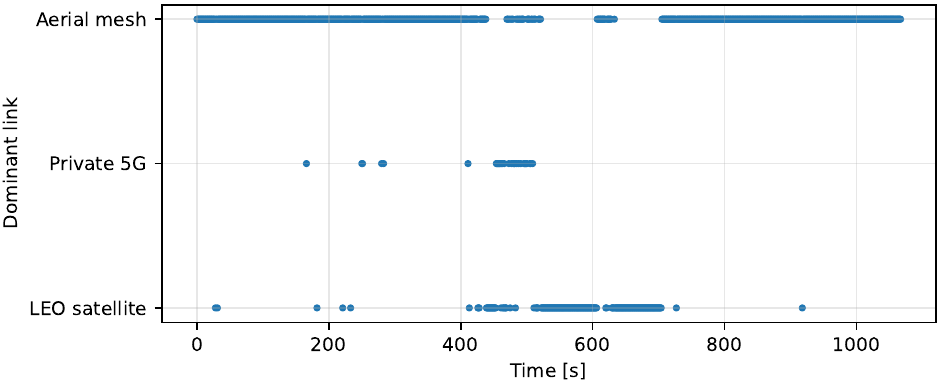}
    \caption{Dominant communication link over time, defined as the link carrying the largest fraction
    of traffic at each instant.}
    \label{fig:link_dominance}
    \vspace{-3mm}
\end{figure}

Figure~\ref{fig:link_dominance} further illustrates the dominant communication link over time, defined as the link carrying the largest fraction of traffic at each instant. The results highlight dynamic link switching behavior as the UAV moves through different coverage regions. Despite frequent changes in link dominance, data delivery continues throughout the experiment, indicating that lossless multipath transport successfully preserves connectivity continuity as long as at least one link remains operational.

\subsection{Achieved Data Rate and Bursty Delivery}
\begin{figure}[t]
    \centering
    \includegraphics[width=\columnwidth]{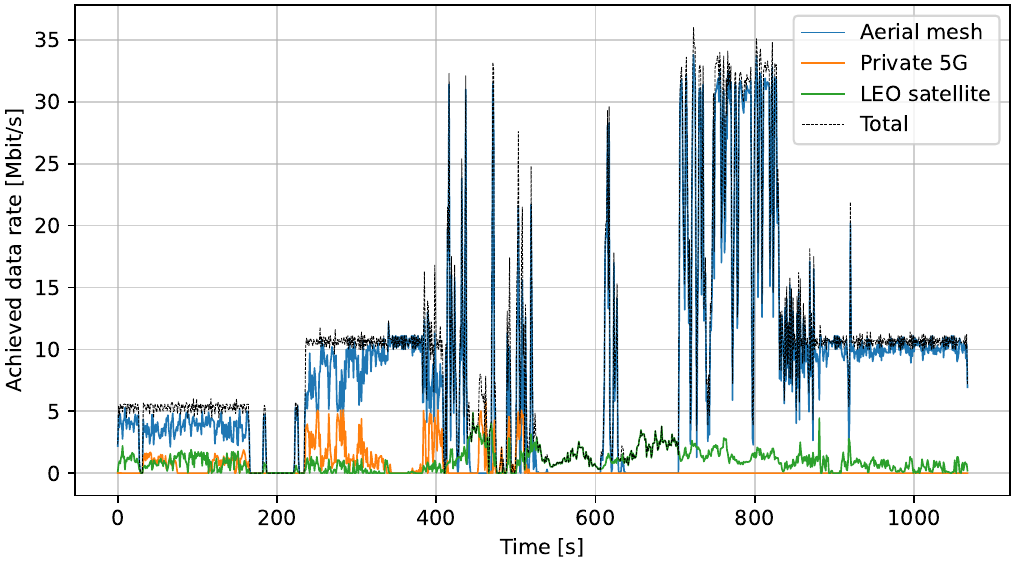}
    \caption{Instantaneous achieved data rate over time during the UAV flight. Periods of reduced
    throughput correspond to buffer accumulation, while short bursts indicate buffer drainage under
    in-order multipath delivery.}
    \label{fig:datarate_distance}
    \vspace{-3mm}
\end{figure}

Figure~\ref{fig:datarate_distance} shows the achieved application-layer data rate as a function of time during the flight. 
The measured data rate varies significantly, ranging from near zero up to approximately $35$ Mbit/s. Periods of near-zero throughput correspond to buffer accumulation, while higher-rate intervals reflect bursty packet release when delayed packets arrive.
These variations are closely correlated with the availability of the aerial mesh link.
When the mesh link is available, the aggregate throughput increases substantially. When it becomes unavailable, the achieved data rate drops toward the capacity limits of the private cellular and satellite links. Importantly, the delivered data rate does not remain constant at the application source rate. Instead, periods of reduced throughput are followed by short intervals of elevated data rates, indicating buffering and bursty packet release caused by in-order delivery semantics under heterogeneous RTTs.

\subsection{Implications for Real-Time Service Continuity}

The results reveal a mismatch between connectivity continuity and service continuity in heterogeneous UAV networks. Although lossless multipath transport successfully maintains end-to-end connectivity throughout the experiment, it fails to sustain continuous low-latency delivery for the $10$~Mbit/s real-time application.

When the aerial mesh link is unavailable, the aggregate capacity of the remaining links is insufficient to carry the application traffic without buffering. Moreover, the large RTT of the satellite link and fluctuations in cellular link quality limit efficient utilization of their nominal capacity. As a result, packets accumulate during low-throughput periods and are delivered in bursts when delayed packets arrive, leading to increased packet age and service interruptions.

These results confirm that maintaining connectivity alone is insufficient to guarantee service continuity for latency-sensitive UAV applications when path delays and capacities are highly heterogeneous.

\section{{{Mechanism~Analysis:~Delay-Induced Buffer Amplification}}}
\label{sec:mechanism}

In our MPTCP setup, the lowest-RTT scheduler prioritizes the path with the smallest RTT to minimize end-to-end latency, while Reno congestion control reduces the congestion window upon loss. In addition, MPTCP enforces lossless, in-order delivery, which can amplify sender- and receiver-side buffering under RTT and loss heterogeneity across paths. The measurement results in Section~\ref{sec:measurements} show that lossless multipath transport preserves end-to-end connectivity under heterogeneous link availability, but induces substantial rate fluctuations and service continuity violations for real-time UAV traffic. In this section, we analyze the underlying transport-layer mechanisms responsible for this behavior. Our goal is not to propose an alternative protocol, but rather to explain why the observed effects arise when conventional lossless, in-order multipath transport is applied to integrated terrestrial--non-terrestrial networks.

\subsection{In-Order Delivery under Heterogeneous RTTs}

Lossless multipath transport protocols enforce reliable and in-order packet delivery at the receiver. 
To preserve in-order delivery due to the heterogeneity of the links, the receiver must buffer early-arriving packets until the missing packets are received.

In the considered multi-technology UAV network, Fig.~\ref{fig:rtt_cdf} shows that RTTs differ by more than an order of magnitude across the available paths. As a result, packets transmitted over the aerial mesh link may arrive significantly earlier than packets transmitted over the terrestrial cellular or LEO satellite links. This delay asymmetry creates persistent packet reordering at the receiver whenever traffic is distributed across multiple paths.

\subsection{Buffer Growth under Constant-Rate Traffic}

Consider a real-time application generating packets at a constant rate $r$. When packets are transmitted concurrently over two paths $i$ and $j$ with RTTs $R_i \ll R_j$, packets arriving via the lower-RTT path accumulate in the receiver buffer while waiting for packets arriving via the higher-RTT path. The buffer occupancy grows over time in proportion to the rate at which early packets arrive and the delay difference between the paths.

This effect is exacerbated when the higher-RTT path also has limited capacity, as shown in Table~\ref{tab:link_performance}. When the application rate exceeds the capacity of the slower paths, packets must be buffered at the sender or dropped if it exceeds the buffer size until sufficient capacity becomes available or delayed packets arrive. Consequently, buffering is not a transient phenomenon but persists over extended periods whenever heterogeneous paths are used concurrently.

\subsection{Burst Release and Rate Fluctuations}

When delayed packets from the high-RTT path finally arrive, the receiver can release a large number of buffered packets in rapid succession to restore in-order delivery. This results in bursty packet delivery at the application interface, even though the source generates traffic at a steady rate.

The achieved data rate fluctuations observed in Fig.~\ref{fig:datarate_distance}, where the instantaneous rate varies between approximately zero and $35$~Mbit/s, are consistent with this mechanism. Periods of reduced throughput correspond to buffer accumulation, while short intervals of elevated throughput correspond to buffer drainage. Importantly, these bursts are induced by transport-layer reordering and buffering, rather than by changes in source behavior or application demand.

\subsection{Impact on Service Continuity}

The buffering and burst release mechanism described above directly affects service continuity for latency-sensitive UAV applications. During buffer accumulation, packet age increases as packets wait for delayed arrivals. During burst release, packets that have already exceeded the application delay bound may be delivered in rapid succession, rendering them unusable for real-time consumption.

Crucially, these service continuity violations occur even though connectivity continuity is preserved throughout the experiment. As long as at least one path remains available, lossless multipath transport can deliver packets eventually, but it cannot guarantee that packets arrive within the delay bounds required by real-time services.

The above analysis leads to a central insight supported by the measurements. 
In multi-technology UAV networks with highly heterogeneous RTTs and capacity constraints, lossless in-order multipath transport inherently trades service continuity for connectivity continuity.

This tradeoff is a direct consequence of transport-layer reliability and ordering semantics, rather than a deficiency of any specific implementation. As such, it motivates the exploration of multipath designs that explicitly account for delay constraints and tolerate controlled packet loss or reordering when supporting real-time UAV services.
\section{Conclusion}
\label{sec:conclusion}

This paper examined the behavior of lossless, in-order multipath transport in multi-technology UAV
networks integrating aerial mesh, terrestrial cellular, and non-terrestrial links. Using measurement
evidence from real UAV flight experiments, we showed that while multipath aggregation can
effectively preserve end-to-end connectivity under link outages, it can fail to support latency-sensitive
UAV services when path delays and capacities are highly heterogeneous.

To interpret these observations, we formally distinguished between \emph{connectivity continuity} and
\emph{service continuity}. Our results demonstrate that maintaining connectivity across heterogeneous
access technologies is a necessary but insufficient condition for meeting real-time service
requirements. In particular, the interaction between in-order delivery semantics, large RTT
disparities, and per-link capacity constraints leads to receiver-side buffering and bursty packet
delivery, causing violations of application delay bounds even when aggregate connectivity is
available.

Rather than proposing new transport mechanisms, this work provides a system-level characterization
and problem formulation that clarifies why conventional lossless multipath abstractions are ill-suited
for strict real-time UAV applications in integrated terrestrial--non-terrestrial environments. These
findings suggest that future multi-technology network designs should explicitly account for
service-level timeliness, rather than relying solely on connectivity-oriented robustness, in order to support safety-critical and mission-critical UAV operations better.

\bibliographystyle{IEEEtran}
\bibliography{main}
\end{document}